\documentclass{ar-1col}
\usepackage[numbers]{natbib}

\jname{Xxxx. Xxx. Xxx. Xxx.}
\jvol{AA}
\jyear{2018}
\doi{10.1146/((please add article doi))}

\def\thingie{\hbox{\kern-9pt\raise1pt%
         \hbox{{\fiverm(}{\lower1.5pt\hbox{\twelvebf--}}{\fiverm)}}}}
\def\pmdiff#1#2{\raise.5ex\hbox{$\sss +#1}$%
    \kern-2.8em\lower1ex\hbox{${\sss-#2}$}} 
\def\barp{{\raise.35ex\hbox{${\sss (}$}}---{\raise.35ex\hbox{${\sss )}$}}}				
\def\lsim{\;\raisebox{-.6ex}{$\stackrel{<}{\sim}$}\;}

\def\ra{\rightarrow}
\def\poptm{\raise.5ex\hbox{${\st +}$}%
    \kern-0.5em\lower.5ex\hbox{${\sss (}{\st -}{\sss )}$} }  
\def\bdbarp{\hbox{$B_d$\kern-1.4em\raise1.4ex\hbox{\barp}}}
\def\nlpbarp{\hbox{$\nu_{\ell^{\prime}}$\kern-1.4em \raise1.4ex\hbox{\barp}}}

\def\ra{\rightarrow}

\newcommand{\Eq}[1]{Eq.~(\ref{eq#1})}
\newcommand{\beq}{\begin{equation}}
\newcommand{\eeq}{\end{equation}}

\begin{document}

\markboth{Balantekin and Kayser}{Dirac and Majorana masses}

\title{On the Properties \newline of Neutrinos}


\author{A. Baha Balantekin,$^1$ Boris Kayser$^2$ 
\affil{$^1$Department of Physics, University of Wisconsin, Madison WI, 53706, USA; email:baha@physics.wisc.edu}
\affil{$^2$Fermi National Accelerator Laboratory, Batavia IL, 60510, USA; email: boris@fnal.gov}}

\begin{abstract}
Our present understanding of neutrino properties is reviewed with a particular emphasis on observable differences between Majorana and  Dirac neutrinos. Current and future experimental efforts towards measuring neutrino properties are summarized. Consequences of the Majorana vs.\ Dirac nature of neutrinos on neutrino masses, neutrino decays, and neutrino electromagnetic properties are described.   
\end{abstract}

\maketitle

\tableofcontents



\section{INTRODUCTION}

In the last several decades, neutrino physics has progressed at a breathtaking pace. We now know that only three active flavors couple to the W and Z. The electroweak eigenstates of these 
neutrinos are linear combinations of their mass eigenstates. Parts of the worldwide neutrino program have reached precision stage. Short- and long-baseline neutrino oscillation experiments, as well as observations of neutrinos produced by the nuclear fusion reactions in the Sun and those produced by cosmic-rays in the upper atmosphere, have determined two non-zero differences between the squares of the masses of the three mass eigenstates, demonstrating that at least two of these eigenstates have non-zero mass. The same experiments have also measured three of the parameters of the mixing transformation, so-called mixing angles, with unprecedented precision. 
Nevertheless, many unanswered questions remain. 
We have only limits on the absolute values of the neutrino masses from direct detection experiments and cosmology. The unitarity of the transformation connecting the mass eigenstates to the electroweak eigenstates is not firmly established. There are tantalizing hints, but no firm evidence, of the existence of sterile neutrinos that do not couple to the vector bosons of the Standard Model, but nevertheless mix with the neutrinos that do. We do not know the transformation properties of the neutrinos under particle-antiparticle conjugation (i.e.\ whether the neutrinos are Majorana or Dirac fermions). 
We do not know just how small the neutrino magnetic moments are. We have just started exploring the full potential of the neutrinos in astrophysics and cosmology. Knowing the correct answer to these questions 
could lead to paradigm-shifting developments in physics and astrophysics. 
Two of the three leptonic mixing angles are much larger than the quark mixing angles, a fact that needs to be understood by an appropriate extension of the Standard Model. CP-violation in the leptonic sector
may shed light on the origin of the baryon-antibaryon asymmetry in the Universe. The interaction of neutrinos with ordinary matter is rather feeble except when the density is very large. Consequently, neutrinos can easily transfer a significant amount of energy and entropy in astrophysical settings, impacting many cosmic phenomena. 

The purpose of this article is to explore our current understanding of the properties of neutrinos. In particular we cover neutrino masses, 
the nature of the relation between neutrinos and antineutrinos (Dirac vs.\ Majorana), and the electromagnetic properties of neutrinos, with particular emphasis on the connection between the latter two subjects. We will limit the discussion of empirical observations primarily to terrestrial probes, and briefly mention relevant insights from astrophysics and cosmology. 

In view of the central importance of the question of whether the neutrinos are Dirac particles or Majorana particles, we conclude this introduction by defining these terms. A Dirac neutrino $\nu^{(D)}$ is one that is distinct from its antiparticle: $\nu^{(D)} \neq \overline{\nu^{D}}$. When the neutrinos are Dirac particles, there is a conserved lepton number $L$ that is $+1$ for leptons, both charged and neutral, and $-1$ for antileptons, both charged and neutral. The distinction between a Dirac neutrino and its antiparticle is then that they carry opposite values of $L$. A Majorana neutrino $\nu^{(M)}$ is one that is identical to its antiparticle: $\nu^{(M)} = \overline{\nu^{M}}$. When the neutrinos are Majorana particles, there is no conserved lepton number \cite{refA}.

The free field of a Dirac neutrino is a spinor with four independent components. This field is distinct from its charge conjugate. In contrast, the free field of a Majorana neutrino is identical to its charge conjugate, apart from a possible phase factor. While this field may be written in four-component form (as it will be in {Sec.~\ref{sec3.2}), only two of its components are independent.

\section{PRESENT EXPERIMENTAL STATUS}

Neutrino oscillation experiments at different baselines have firmly established that the neutrino flavor states that are produced by the weak interactions are combinations of mass eigenstates, i.e.,
\begin{equation}
| \nu_f \rangle = \sum_i U_{fi} |\nu_i \rangle  \;\; ,
\end{equation}
where $f$ and $i$ are flavor and mass basis indices, respectively. 
Precise measurement of the invisible decay width of the $Z$ boson restricts the number of flavors that can participate in weak interactions to three so-called active neutrinos: $f= e, \mu, \tau$. Clearly, to have three linearly-independent flavor eigenstates, one needs at least three mass eigenstates. If the number of mass eigenstates is also three, imposing the condition that one can use either flavor or mass basis to describe the same physics requires the $3 \times 3$ matrix $U$ to be unitary. However, there is no fundamental reason  or symmetry principle that would limit the number of mass eigenstates to three. In case of more than three mass eigenstates, the only constraint is that only three ``active'' combinations of these mass eigenstates couple to the electroweak gauge bosons; the remaining orthogonal ``sterile'' combinations do not. Here one  point is worth clarifying: Sometimes in the literature mass eigenstates are called sterile states if their contributions to the three active flavors are very small.  (For example there may be a fourth mass eigenstate and the coefficient of that mass eigenstate in the linear combination that defines the electron neutrino is likely to be very small.) Strictly speaking, such a description is misleading since the word ``sterile'' refers to the lack of interaction with electroweak gauge bosons, hence should be reserved for flavor states. 

A $3 \times 3$ unitary matrix with unit determinant has 9 independent variables. This matrix can be parameterized using trigonometric functions of three Euler angles and six additional phases. For Dirac neutrinos five of these phases can be absorbed in the definitions of neutrino states and one is left with three angles and one CP-violating phase describing mixing of three flavors. However, for Majorana neutrinos it is not possible to absorb two more of these additional phases since the Majorana fields must remain self charge-conjugate.
The number of parameters quickly increases with increasing number of mass eigenstates. For example, inclusion of a fourth mass eigenstate necessitates a parameterization with six angles and three CP-violating phases to describe mixing of three active and one sterile Dirac flavor states. 

There are a number of anomalies in various experiments that can be interpreted as coming from one or more sterile neutrino admixtures. However, concrete experimental evidence for sterile neutrinos is still lacking. For three flavors, the combination of the solar, atmospheric, reactor and accelerator experiments have measured the three angles in the mixing matrix with a different precision for each angle. The value of the CP-violating phase is not yet determined.

\subsection{Oscillation experiments}

From the equation describing the evolution of mass eigenstates for non-interacting neutrinos, 
\begin{equation}
i \frac{\partial}{\partial t} | \nu_i \rangle = E_i  | \nu_i \rangle \;\; ,
\end{equation}
one can write down an equation describing evolution in the flavor space:
\begin{equation}
\label{free_neutrinos}
i \frac{\partial}{\partial t} | \nu_f \rangle = \sum_{f'}  \left( U \Lambda U^{\dagger} \right)_{f f'} | \nu_{f'} \rangle , \>\>\>\>  \Lambda_{ij} = E_i \delta_{ij}  \;\; .
\end{equation}
From this equation it is clear that flavor changes will depend on the differences $E_i-E_j \sim \frac{m_i^2 - m_j^2}{2E}$, since one can write 
$\Lambda$ as the sum of a matrix proportional to the identity and a matrix that depends only on those differences. Hence, oscillation experiments where neutrinos do not interact between production and detection only measure the differences $\delta m^2_{ij} = m_i^2 - m_j^2$, not the individual masses. In fact, experiments looking only at the disappearance of the original flavor are not even sensitive to the signs of these differences:
\begin{equation}
P(\nu_f \to \nu_f) = 1 - 2 \sum_{i \neq j} |U_{fi}|^2 |U_{fj}|^2 \sin^2 \left( \frac{\delta m^2_{ij}}{4E} L \right) \;\; .
\end{equation}

If neutrinos interact between their source and their detection the situation changes. Except in circumstances where matter densities are exceedingly large, their collisions with background particles can be neglected with reasonably good accuracy since the relevant cross sections are proportional to $G_F^2$, where $G_F$ is the Fermi coupling constant). However, neutrinos would coherently scatter in the forward direction from the background particles with an amplitude proportional to $G_F$. Including this effect modifies Eq.~(\ref{free_neutrinos}) according to
\begin{equation}
 \left( U \Lambda U^{\dagger} \right)_{f f'} \to  \left( U \Lambda U^{\dagger} \right)_{f f'} + a_f \delta_{f f'} 
\end{equation}
if the background is static and free of large polarizing magnetic fields. In the Standard Model both the charged- and neutral-current interactions contribute to the quantities $a_f$. At the tree level (but not when one includes the first-order loop corrections), the neutral current contributes the same amount to all active flavors and, in the absence of sterile neutrinos, the resulting proportional-to-identity matrix does not impact the flavor change.  The remaining term, $a_e$, is proportional to the background electron density in the Standard Model for locally charge-neutral backgrounds. The value of the combined $\delta m^2$ and $a_e$ terms can vary significantly depending on the sign of $\delta m^2$; it can even be zero at the so-called Mikheev, Smirnov, Wolfenstein (MSW) resonance \cite{Mikheev:1986gs,Wolfenstein:1979ni}. 

For three active flavors the mixing matrix can be parameterized as
\begin{equation}
U = 
\left[
\begin{array}{ccc}
 c_{12}c_{13} &  s_{12}c_{13} & s_{13}e^{-i\delta}  \\
-s_{12}c_{23}  -c_{12}s_{23}s_{13}e^{i \delta}  & c_{12}c_{23}  -s_{12}s_{23}s_{13}e^{i \delta}  & s_{23}c_{13}  \\
s_{12}s_{23}  -c_{12}c_{23}s_{13}e^{i \delta}   & -c_{12}s_{23}  -s_{12}c_{23}s_{13}e^{i \delta}   &   c_{23}c_{13}
\end{array}
\right] \;\; ,
\end{equation}
where we used the notation $c_{ij}= \cos \theta_{ij}$, $s_{ij} = \sin \theta_{ij}$ with all angles in the range $0 \le \theta_{ij} \le \pi/2$ and designated the CP-violating Dirac phase as $\delta$. Majorana phases, which do not appear in the probabilities measured by the 
oscillation experiments, are not shown. 
The Euler angles are measured to be $\sin^2 \theta_{12} = 0.307\pm 0.013$, $\sin^2 \theta_{23} = 0.51 \pm 0.04$, and 
$\sin^2 \theta_{13} = 0.0210 \pm 0.0011$ \cite{Patrignani:2016xqp}. 
For three flavors one can write down two distinct differences of the squares of masses. Combining all the measurements gives the smaller one as $(7.53 \pm 0.18) \times 10^{-5}$ eV$^2$ and the larger one as $\sim 2 \times 10^{-3}$ eV$^2$ \cite{Patrignani:2016xqp}. Solar neutrino physics has determined that, of the two mass eigenstates separated by the smaller difference, $\delta m^2_{21}$, the mass eigenstate that is $\sim 2/3$ of electron flavor is the lighter one.
 For the larger difference, $\delta m^2_{31}$, there remain two possibilities: The possibility of $\delta m^2_{31} > 0$ is referred to as a ``normal'' hierarchy and the possibility $\delta m^2_{31}<0 $ is referred to as an ``inverted'' hierarchy. Assuming a normal hierarchy yields a value of $(2.45 \pm 0.05) \times 10^{-3}$ eV$^2$ for the larger $\delta m^2$. For the inverted mass hierarchy, one obtains  a value of $(2.52 \pm 0.05) \times 10^{-3}$ eV$^2$. 

Experiments measuring the appearance of a flavor not present at the neutrino source are sensitive to the CP-violating phase, and the sign of $\delta m^2$. Note that oscillation experiments do not determine the overall neutrino mass scale, i.e.\ the value of the smallest neutrino mass.

\subsection{Direct neutrino mass measurements}

It is possible to measure the neutrino masses using nuclear beta decays. Near the endpoint of the beta spectrum, corresponding to the highest values of the measured electron energies, at least two of the mass eigenstates are nonrelativistic, which implies a linear dependence of the decay probability on the masses. The maximum kinetic energy of the electron is $Q=E_0 - m_e$ where $E_0$ is the total decay energy. In a beta decay experiment, the spectrum is measured up to an electron energy $E$ near $E_0$. The fraction of decays in the interval $E_0-E$ is given by 
$(E_0-E)^3/Q^3$. Hence one needs a nucleus  with a small value of the $Q$. A relatively short decay lifetime is also helpful to reduce the amount of line broadening. The tritium nucleus satisfies both of these constraints. 

Direct mass measurements are very robust since they only depend on conservation of energy. Since neutrinos mix, these measurements probe the quantity \cite{refFS1}
\begin{equation}
m^2_{\beta} = \sum_i |U_{ei}|^2 m_i^2  \;\; . 
\end{equation}
So far, two experiments carefully measured the endpoint of the tritium beta-decay spectrum. The Troitsk experiment reported $m_{\beta}^{2} =-0.67 \pm 2.53$ eV$^2$, corresponding to a limit of  $m_\beta < 2.2$ eV \cite{Aseev:2011dq}. 
The Mainz experiments reported $m^2_{\beta} = - 0.6 \pm 2.2 ({\rm stat}) \pm 2.1({\rm syst})$ eV$^2$, corresponding to a limit of  $m_\beta < 2.3$ eV \cite{Kraus:2004zw}.  A third experiment, KATRIN, is a high resolution spectrometer based on magnetic adiabatic collimation combined with an electrostatic filter, using a well-characterized gaseous tritium source. It is expected to start taking data in 2018 and eventually reach a sensitivity of 0.2 eV \cite{Osipowicz:2001sq}. To increase the sensitivity of a KATRIN-like experiment one needs a larger spectrometer. However, given the size of the existing KATRIN spectrometer, scaling up this approach does not seem to be realistic. To circumvent this problem another measurement based on cyclotron radiation emission spectroscopy has been proposed. This approach uses the principle that the energy of the emitted electron can be determined very accurately by detecting the radiation it emits when moving in a magnetic field. The planned  experiment, Project 8, has the potential to reach sensitivities down to $m_\beta  \sim 40$ meV using an atomic tritium source 
\cite{Esfahani:2017dmu}.

\subsection{Cosmological considerations}

A quantity that is sometimes misstated as the number of neutrino species is the quantity called $N_{\rm eff}$ in the Big Bang cosmology. 
This quantity is defined in terms of the radiation energy density deduced from the observations of cosmic microwave background radiation 
at photon decoupling as 
\begin{equation}
\rho_{\rm rad} \equiv \frac{\pi^2}{15} T^4_{\gamma} \left[ 1 + \frac{7}{8} N_{\rm eff} \left( \frac{4}{11} \right)^{4/3} \right], 
\end{equation}
where $T_{\gamma}$ is the photon temperature. The radiation density on the left side of this equation receives contributions from photons, three active flavors of neutrinos as well as antineutrinos and all other particles that may be present. In the limit all masses and lepton asymmetries are set to zero, all interactions and plasma effects are ignored, and no other particles are assumed to be present besides photons and active neutrinos, $N_{\rm eff}$ takes the value of $3$. Including the Standard Model interactions and masses slightly changes the thermal blackbody spectra of neutrinos and increases this value by a small amount. Planck Collaboration reports a value of $ N_ {\rm eff} = 3.15 \pm 0.23$ \cite{Ade:2015xua}. 
A careful evaluation of the assumptions that need to be taken into account to assess what $N_{\rm eff}$ represents is given in Ref.~\cite{Grohs:2014rea}. 

Although the value of the Hubble parameter deduced from the observations of the cosmic microwave radiation and that deduced from measurements of distances of galaxies do not quite agree, accelerated expansion of the Universe is a widely accepted conclusion. Similarly, standard cosmology predicts the existence of a neutrino background left over from the Early Universe with a temperature of $1.9^o$ K. If the sum of the masses of the neutrinos exceeds a certain value this expansion can be halted. Note that this argument provides an upper limit to $\sum_i m_i$.  The Planck Collaboration reports an upper limit of 0.23 eV for this sum  \cite{Ade:2015xua}. 

From oscillation experiments we know that at least two neutrinos have a non-zero mass. The lightest neutrino may also be massive, but it could also have zero mass. This leaves open the possibility where one of the cosmic background neutrinos  is relativistic, but the other two are non-relativistic. We discuss in Sec.~\ref{sec3.5} an interesting consequence of such a possibility.

\section{DIRAC AND MAJORANA MASSES AND THEIR CONSEQUENCES}

\subsection{Neutrino mass}
The discovery and study of the Higgs boson at CERN's Large Hadron Collider has provided strong evidence that the quarks and charged leptons derive their masses from an interaction with the Standard Model Higgs field. Conceivably, the neutrinos derive their masses in the same way. However, because the neutrinos are electrically neutral, the origin of their masses could involve an ingredient---a ``Majorana mass''---that is forbidden to the quarks and charged leptons.

The essence of a Majorana mass, and how this mass differs from a ``Dirac mass'', which is the kind of mass a quark has, is depicted in \textbf{Figure \ref{figK.1}}. For simplicity, this figure treats a world with just one flavor, and, correspondingly, just one neutrino mass eigenstate. The neutrinos $\nu$ and $\bar{\nu}$  in the figure are not the mass eigenstate and its antiparticle, but underlying neutrino states in terms of which we construct the picture of neutrino mass. These underlying states $\nu$ and $\bar{\nu}$ are distinct from each other.
\begin{figure}[h]
\includegraphics[width=5in]{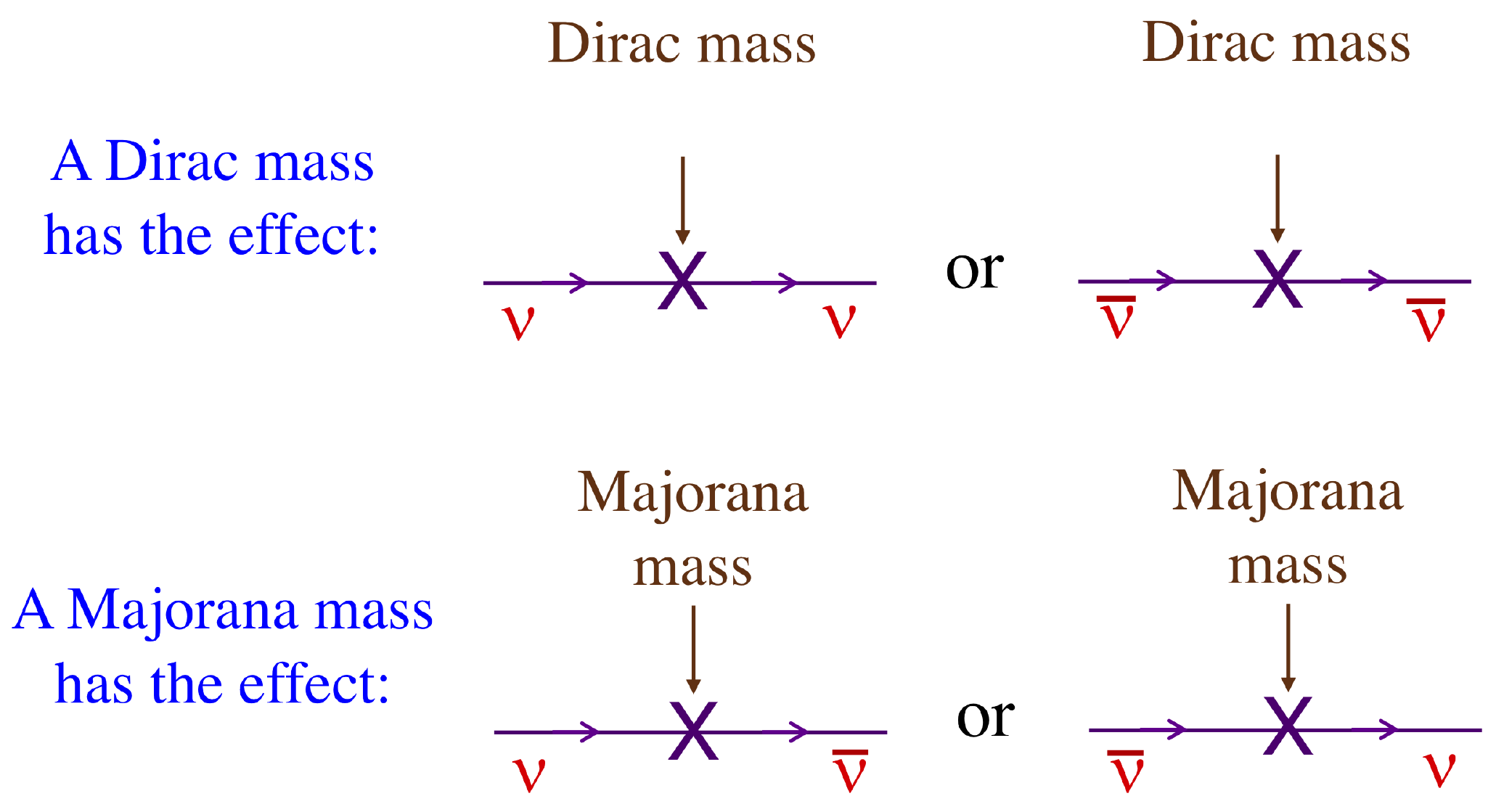}
\caption{The effects of Dirac and Majorana mass terms in the Lagrangian.}
\label{figK.1}
\end{figure}

As shown in \textbf{Figure \ref{figK.1}}, when a Dirac mass term in the Lagrangian acts on an incoming $\nu$, it leaves this particle a $\nu$, and when it acts on a $\bar{\nu}$, it leaves this particle a $\bar{\nu}$. In constrast, when a Majorana mass term acts on a $\nu$, it turns it into a $\bar{\nu}$, and when it acts on a $\bar{\nu}$, it turns it into a $\nu$. Thus, Majorana neutrino masses do not conserve the lepton number $L$ that is defined as +1 for a lepton, neutral or charged, and --1 for an antilepton.

Beyond neutrinos, a Majorana mass acting on any fermion turns it into its antiparticle. If the fermion is electrically charged, this transition reverses its electric charge, violating electric-charge conservation. That is why the quarks and charged leptons cannot have Majorana masses.

When the neutrino mass term is a Majorana mass term, its mass eigenstate is $\nu + \bar{\nu}$, since this is clearly the state that the mass term sends back into itself, as depicted in \textbf{Figure \ref{figK.2}}.
\begin{figure}[h]
\includegraphics[width=2.5in]{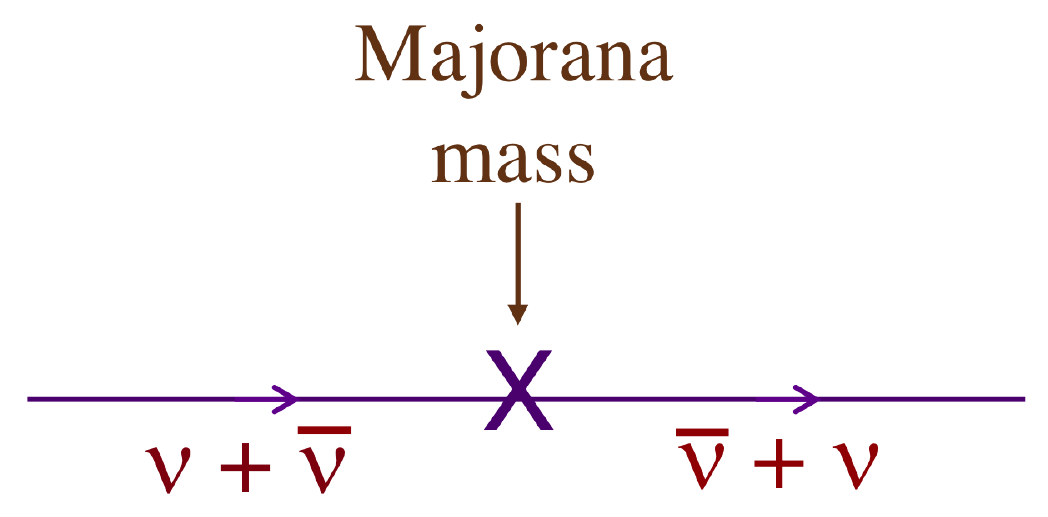}
\caption{Effect of a Majorana mass term on its mass eigenstate, $\nu + \bar{\nu}$.}
\label{figK.2}
\end{figure}
Noting that $\nu + \bar{\nu}$ is self-conjugate under particle-antiparticle interchange, we see that when the neutrino mass term is a Majorana mass term, the neutrino mass eigenstate will be a Majorana neutrino. It is easily shown that when there are several flavors, and correspondingly several neutrino mass eigenstates, if the neutrino mass term---now a matrix in flavor space---is a Majorana mass term, all the neutrino mass eigenstates will be Majorana neutrinos. Correspondingly, if the neutrino mass term is a Dirac mass term, all the mass eigenstates will be Dirac neutrinos. 

What if, in the real world of multiple, mixed flavors, there are both Dirac and Majorana neutrino mass terms? So long as there are Majorana mass terms, the lepton number $L$ that would distinguish Dirac neutrinos from their antiparticles is no longer conserved. Thus, one would expect that the neutrino mass eigenstates will be Majorana particles. This expectation is indeed correct, as shown in  \cite{refK.1}. If there are $n$ flavors, and one starts with Dirac mass terms alone, there will be $n$ Dirac mass eigenstates. Of course, each mass eigenstate will be a collection of four states: the two helicity states of a neutrino, plus the two of an antineutrino. If one then adds Majorana mass terms, the $n$ Dirac neutrinos will be replaced by $2n$ Majorana neutrinos, each comprising just the two helicity states of any spin-$1/2$ particle.

Let us now turn to the possible origins of Dirac and Majorana neutrino masses. For simplicity, we shall neglect mixing. Any mass term couples two neutrino fields of opposite chirality to each other. Charge conjugating a field of definite chirality reverses its chirality, and in a Majorana mass term, one of the two coupled neutrino fields is simply the charge conjugate of the other. In a Dirac mass term, the chirally left-handed neutrino field $\nu_L$ belongs to a Standard Model (SM) weak-isospin doublet, while the chirally right-handed one $\nu_R$ must be added to the SM, which contains no right-handed neutrino fields, before a Dirac mass term is possible. Once $\nu_R$ has been introduced, the Lagrangian may contain the Yukawa interaction
\beq
{\cal L}_Y = - y \, H^0 \, \overline{\nu_R} \, \nu_L + h.c. \; \;,
\label{eqK.1}
\eeq
where $H^0$ is the SM neutral Higgs boson field, and $y$ is a real Yukawa coupling constant. When $H^0$ develops its vacuum expectation value $\langle H^0 \rangle _0 = 174$ GeV $\equiv v$, this Yukawa interaction leads to the Dirac mass term
\beq
{\cal L}_D = - m_D  \, \overline{\nu_R} \, \nu_L + h.c. \;\; ,
\label{eqK.2}
\eeq
where $m_D = y v$ is the Dirac neutrino mass. If this mass term is the sole source of mass for a neutrino $\nu$, then the field of this neutrino is given by $\nu = \nu_L + \nu_R$. In terms of this field, the mass term can be written as
\beq
{\cal L}_D = - m_D  \, \bar{\nu} \, \nu \; \; ,
\label{eqK.3}
\eeq
a form from which it is obvious that $m_D$ is the mass of $\nu$. If $\nu$ is one of the three established neutrino mass eigenstates, whose masses are $\lsim 0.1$ eV, then $y = m_D / v \lsim 10^{-12}$. A coupling constant this much smaller than unity leaves many theorists skeptical of the notion that Dirac mass terms are the sole source of neutrino masses.

Majorana mass terms can have several different origins, all of which entail physics that is far outside the SM. Once a right-handed field $\nu_R$ has been introduced, we can have a ``right-handed Majorana mass term''
\beq
{\cal L}_R = -\frac{m_R}{2}  \, \overline{(\nu_R)^c} \, \nu_R + h.c. \;\; ,
\label{eqK.4}
\eeq
where the superscript $c$ denotes charge conjugation, and $m_R$ is a positive, real constant. Note that $\overline{(\nu_R)^c} \,\nu_R$ absorbs a neutrino and emits an antineutrino, 
while its Hermitean conjugate, $\overline{\nu_R}\, (\nu_R)^c$, absorbs an antineutrino and emits a neutrino, in conformity with the effects attributed to Majorana mass terms in our earlier discussion. 
Since a right-handed neutrino does not carry any non-zero quantum number that is conserved in the SM, or as far as we know in nature, the right-handed Majorana mass term of \Eq{K.4} does not violate any known conservation law, despite the fact that it transforms neutrinos and antineutrinos into one another. This Majorana mass term does not arise from a neutrino coupling to the SM Higgs field, so its origin is quite different from the Brout-Englert-Higgs mechanism that leads to masses in the SM. If this mass term is the only source of mass for a neutrino $\nu$, than the field of this neutrino is given by $\nu = \nu_R + (\nu_R)^c$. As we see, $\nu$ is a self-conjugate (i.e., Majorana) neutrino, in agreement with our earlier conclusion that when the mass term is a Majorana mass term, its mass eigenstate will be a Majorana neutrino. In terms of the field $\nu$, the right-handed Majorana mass term can be written as
\beq
{\cal L}_R = - \frac{m_R}{2}  \, \bar{\nu} \, \nu + h.c. \;\; .
\label{eqK.5}
\eeq
This has the expected form for the mass term of a mass eigenstate $\nu$ of mass $m_R$, except for the factor of $1/2$. To understand that factor, recall that the mass term for a fermion mass eigenstate must absorb and then reemit that fermion with an amplitude that is the fermion's mass. As we have noted, in the mass term of \Eq{K.5}, the field is a ``Majorana'' (i.e., self-conjugate) field. As a result, in contrast to the mass term for a Dirac neutrino, where only the field $\nu$ can absorb the incoming neutrino, and only the field $\bar{\nu}$ can emit the outgoing one, now there is an additional term in which the field $\bar{\nu}$ absorbs the incoming neutrino, and the field $\nu$ emits the outgoing one. This additional term is equal in size to the sole term in the Dirac case, so the amplitude produced by $(m_R /2) \bar{\nu} \nu$ is $m_R$, not $(m_R / 2)$. Since by definition this amplitude must be the mass of $\nu$, our notation has been so chosen that the parameter $m_R$ is the mass of $\nu$.

Even if there is no right-handed neutrino field, if there is a non-SM weak-isospin triplet of scalar fields $\Delta$, there can be an interaction of the form $\Delta^0 \overline{(\nu_L)^c} \,\nu_L$, where $\Delta^0$ is the neutral member of the triplet. If $\Delta^0$ has a non-zero vacuum expectation value, this interaction leads to a ``left-handed Majorana mass term''
\beq
{\cal L}_L = -\frac{m_L}{2}  \, \overline{(\nu_L)^c} \, \nu_L + h.c. \; \;,
\label{eqK.6}
\eeq
where $m_L$ is a positive, real constant. As in the case of the right-handed Majorana mass term, the eigenstate of this left-handed one is a Majorana particle. Its field is the self-conjugate  $\nu = \nu_L + (\nu_L)^c$.

Non-SM physics at a high mass scale can lead at present-day energies to an effective interaction of the form $\overline{(\nu_L)^c}\,H^0 H^0 \,
\nu_L / \Lambda$, where $\Lambda$ is the high mass scale from whose physics this effective interaction comes \cite{refK.2}. Once the SM Higgs Field $H^0$ develops its vacuum expectation value $v$, this interaction leads in turn to the effective left-handed Majorana mass term
\beq
{\cal L}_L^{\:\mathrm{effective}} = -\frac{m^\prime_L}{2}  \, \overline{(\nu_L)^c} \, \nu_L + h.c. \;\; ,
\label{eqK.7}
\eeq
where $m_L^\prime / 2 = v^2 / \Lambda$.

Perhaps the leading candidate for an explanation of how the neutrino masses, although non-zero, can be so small is the See-Saw mechanism \cite{refK.3}. To explain this mechanism \cite{refK.6}, let us again treat a world with just one flavor. The most straightforward (so-called type-I) See-Saw model adds to the SM of this world a right-handed neutrino $\nu_R$ and both Dirac and right-handed Majorana neutrino mass terms. The neutrino-mass part of the Lagrangian is then
\beq
{\cal L}_m = - m_D\, \overline{\nu_R} \nu_L - \frac{m_R}{2} \overline{(\nu_R)^c} \,\nu_R + h.c.\;\;  ,
\label{eqK.8}
\eeq
where $\nu_L$ is the neutrino in a SM left-handed lepton doublet, and of course $m_D$ and $m_R$ are constants. Using the identity $\overline{(\nu_L)^c} \,m_D\, (\nu_R)^c = \overline{\nu_R} \,m_D \nu_L$, this ${\cal L}_m$ can be rewritten as
\beq
{\cal L}_m = -\frac{1}{2} \left[ \overline{(\nu_L)^c}, \overline{\nu_R} \right] \, 	
	\left[   \begin{array}{cc}
		0 & m_D \\  m_D  & m_R
		\end{array}  \right]
	\left[	\begin{array}{c}
		\nu_L  \\   (\nu_R)^c
		\end{array}   \right]      + h.c.\;\; .
\label{eqK.9}
\eeq
As already discussed, the right-handed Majorana mass term in Eqs. \ref{eqK.8} and \ref{eqK.9} does not violate any known conservation law. Thus, $m_R$ could be extremely large, and the See-Saw model assumes that it is. On the other hand, the Dirac neutrino mass $m_D$ is presumably of the same order as the masses of the quarks and charged leptons, all of which are Dirac masses. Thus, $m_D \ll m_R$.
The neutrino mass matrix
\beq
M_\nu = \left[   \begin{array}{cc}
		0 & m_D \\  m_D  & m_R
		\end{array}  \right]
\label{eqK.10}
\eeq
in \Eq{K.9} can be diagonalized by the transformation
\beq
Z^T M_\nu Z = D_\nu  \;,
\label{eqK.11}
\eeq
where, with the assumption that $m_D/m_R \equiv \rho \ll 1$,
\beq
Z \cong \left[	\begin{array}{cc}
			1 & \rho \\ -\rho & 1
			\end{array}	\right]
			\left[  \begin{array}{cc}
			i & 0  \\ 0 & 1
			\end{array}	\right]  \;,
\label{eqK.12}
\eeq
and
\beq
D_\nu \cong \left[	\begin{array}{cc}
			m_D^2 / m_R & 0  \\  0 & m_R
			\end{array}	\right]	\; .
\label{eqK.13}
\eeq
Here, the second matrix in \Eq{K.12} is included so that both diagonal elements in $D_\nu$ will be positive. Defining
\beq
\left[	\begin{array}{c}		\nu_L^\prime \\  N_L^\prime		\end{array} \right]
\equiv Z^{-1}  \left[ \begin{array}{c}		\nu_L \\  (\nu_R)^c	\end{array} \right] \; \; ,
\label{eqK14}
\eeq
and
\beq
\left[ \begin{array}{c}		\nu \\  N		\end{array} \right]
\equiv  \left[ \begin{array}{c}	\nu _L^\prime + (\nu_L^\prime)^c  \\  
			N_L^\prime + (N_L^\prime)^c		\end{array} \right] \; ,
\label{eqK.15}
\eeq
we may rewrite \Eq{K.8} as
\beq
{\cal L}_m = -\frac{1}{2} \frac{m_D^2}{m_R} \bar{\nu} \nu - \frac{1}{2} m_R \bar{N} N  \;\; .
\label{eqK.16}
\eeq
From this relation, we see that $\nu$ and $N$ are mass eigenstates, and from the definition of \Eq{K.15}, we see that they are Majorana (self-conjugate) particles. \Eq{K.16} shows that the mass of $\nu$ is $m_D^2 / m_R$, while that of $N$ is $m_R$. Thus,
\beq
( \mathrm{Mass \,of \, \nu)} \times (\mathrm{Mass \,of \,}N) = m_D^2 \sim \mathrm{(Mass \,of \,quark \,or \,charged \,lepton)}^2 \; .
\label{eqK.17}
\eeq
This is the famous See-Saw Relation, which states that the heavier $N$ is, the lighter $\nu$ will be.

It is no surprise that the See-Saw model predicts that the neutrino mass eigenstates will be Majorana particles. As already mentioned, any model that includes Majorana masses will do that. Nor is it a surprise that, even though we were treating a world with just one flavor, we ended up with {\em two} Majorana neutrinos. As already mentioned, when a Majorana mass term is added to a Dirac one in a world with $n$ flavors, the $n$ Dirac neutrinos of that world are replaced by $2n$ Majorana neutrinos.

How large is the mass $m_N$ of the heavy neutrino $N$ predicted to be? While there is no sharp prediction because of the unknown parameter $m_D$, if we assume that $m_D$ is of the order of the mass $m_\mu$ of the muon, the charged lepton in the ``middle'' generation of leptons, and take the mass $m_\nu$ of the light neutrino $\nu$ to be $\sim$0.1 eV, as suggested by neutrino oscillation experiments, then the See-Saw Relation predicts that 
\beq
m_N = \frac{m_D^2}{m_\nu} \sim \frac{m_\mu^2}{0.1 \,\mathrm{ eV}} = 10^8 \,\mathrm{GeV} \;\;.
\label{eqK.18}
\eeq
Thus, according to the See-Saw model, the known light neutrinos are a window into physics at a very high mass scale (unless $m_D$ is much smaller than we have guessed). On the other hand, since $10^8$ GeV is obviously far out of reach of current or near future particle accelerators, it may be a while before the See-Saw picture can be tested.

\subsection{Why determining whether neutrinos are Dirac or Majorana particles is very challenging}
\label{sec3.2}

Why is it that we do not know whether neutrinos are Dirac or Majorana particles? To answer this question, we note first that all the neutrinos we have so far been able to study directly have been ultra-relativistic. As we shall explain, when a neutrino is ultra-relativistic, its behavior is almost completely insensitive, under almost all circumstances, to whether it is a Dirac particle or a Majorana one. 

There may or may not be a conserved lepton number $L$ that distinguishes antileptons from leptons. However, regardless of whether such a conserved quantum number exists, the SM weak interactions are chirally left-handed. 
As a result, when the particle we call the ``neutrino'', $\nu$, is created in, for example, the decay $W^+ \ra e^+ + \nu$, this ``$\nu$'' will be of left-handed (i.e., negative) helicity extremely close to 100\% of the time. In contrast, when the particle we call the ``antineutrino'', $\bar{\nu}$, is created in $W^- \ra e^- + \bar{\nu}$, this ``$\bar{\nu}$'' will be of right-handed (i.e., positive) helicity extremely close to 100\% of the time. In the Majorana case, helicity will be the sole difference between this  ``$\nu$'' and this ``$\bar{\nu}$'' .

Now, suppose the ``$\nu$'' or the ``$\bar{\nu}$'' that has been created in $W$ boson decay interacts with some target via the SM charged-current weak interaction, creating a charged lepton in the process. Neglecting mixing, the leptonic part of the first-generation weak-interaction Lagrangian is
\beq
{\cal L}_{cc} = -\frac{g}{\sqrt{2}} \left[ \bar{e} \gamma^\lambda \frac{(1-\gamma_5)}{2} \nu W^-_\lambda + \bar{\nu}  \gamma^\lambda \frac{(1-\gamma_5)}{2} e W^+_\lambda \right] \;\; ,
\label{eqK.19}
\eeq
where $g$ is the semiweak coupling constant.
If neutrinos are Dirac particles, the lepton number $L$ is conserved, so the $\nu$ created in $W^+$ decay, with $L=+1$, can create only an $e^-$, not an $e^+$, and it will do this via the first term on the right-hand side of \Eq{K.19}. Similarly, the $\bar{\nu}$ created in $W^-$ decay, with $L=-1$, can create only an $e^+$, not an $e^-$, and it will do that via the second term in \Eq{K.19}.

Now suppose that neutrinos are Majorana particles. Then there is no longer a conserved lepton number, and the neutrino field $\nu$ is now a Majorana field that, with a suitable choice of phase convention, takes the form
\beq
\nu = \int \frac{d^3p}{(2\pi)^3} \frac{1}{\sqrt{2E_p}} \sum_h (f_{\vec{p},h} u_{\vec{p},h} e^{-ipx} + f_{\vec{p},h}^\dag v_{\vec{p},h} e^{ipx})  \; \;.
\label{eqK.20}
\eeq
Here, $\vec{p}$ is the momentum of the neutrino, $E_p$ is its energy, and $h$ is its helicity. The operator $f_{\vec{p},h}$ absorbs a neutrino of momentum $\vec{p}$ and helicity $h$, and $f_{\vec{p},h}^\dag $creates such a neutrino. The functions $u_{\vec{p},h}$ and $v_{\vec{p},h}$ are the usual Dirac $u$ and $v$ wave functions for a particle of momentum $\vec{p}$ and helicity $h$. We note that the field operator $\nu$ can both absorb and create a neutrino, and the same is true of 
\beq
\bar{\nu} = \int \frac{d^3p}{(2\pi)^3} \frac{1}{\sqrt{2E_p}} \sum_h (f_{\vec{p},h} \bar{v}_{\vec{p},h} e^{-ipx} + f_{\vec{p},h}^\dag \bar{u}_{\vec{p},h} e^{ipx})  \;\; .
\label{eqK.21}
\eeq
In particular, an incoming neutrino can be absorbed by either the field $\nu$ or the field $\bar{\nu}$. Thus, a Majorana neutrino created in, say, $W^+$ decay can be absorbed, in principle, by either the first or the second term in the charged-current Lagrangian ${\cal L}_{cc}$ of \Eq{K.19}. However, the Majorana neutrino from decay of a $W^+$ will be essentially 100\% polarized with left-handed helicity. Thus, because of the left-handed chirality projection operator $(1-\gamma_5)/2$ in ${\cal L}_{cc}$, and the ultra-relativistic energy of the neutrino, in practice it can be absorbed only by the first term. 
The action of $(1-\gamma_5)/2$ on the $\bar{\nu}$ field in the second term almost totally suppresses this term for a left-handed ultra-relativistic neutrino. Since the action of the first term in  ${\cal L}_{cc}$ always creates an $e^-$, never an $e^+$, our Majorana neutrino form $W^+$ decay will always create an $e^-$ rather than an $e^+$, just as would a Dirac neutrino. In a similar way, a Majorana neutrino born in $W^-$ decay will essentially always be of right-handed helicity, and consequently can in practice be absorbed only by the second term in  ${\cal L}_{cc}$, whose action always produces an $e^+$, never an $e^-$. This is exactly how a Dirac antineutrino would behave.

As the neutrinos from $W$ decay illustrate, for ultra-relativistic neutrinos, helicity is a substitute for lepton number. Even if there is no conserved lepton number, ultra-relativistic neutrinos will behave as if there is such a quantum number. That is, Majorana neutrinos will behave as if they are Dirac neutrinos. To address the question experimentally of whether neutrinos are actually Dirac or actually Majorana particles, we have to find exceptions to this rule, or else find an effective way of working with {\em non-relativistic} neutrinos, or else find some process that addresses the question even though it does not involve neutrinos at all.

\subsection{Double beta decay}

Numerous experiments are seeking to answer the Dirac vs.\ Majorana question through the study of special nuclear decays. 

In the nuclear Shell Model each nucleon is assumed to interact with an average mean field, filling well-defined nuclear shells. Since a heavy nucleus typically has more neutrons than protons, the last filled shell orbital can be very different for protons and neutrons. In a heavy nucleus with an even number of protons and an even number of neutrons, which we denote as X,  like nucleons tend to pair up. If one of the neutrons in  X is replaced by a proton, that proton and the second neutron from the broken pair cannot pair up since they would sit in different shells. Without the benefit of pairing energy this could raise the ground state energy of the new nucleus, which we call Y, to a value higher than the ground state energy of the original nucleus, X. Thus the ground-state to ground-state beta decay of X into Y is energetically impossible. Replacing the odd neutron in Y  
with a second proton could create a third nucleus, Z, the ground state energy of which is lower than the other two nuclei since those two protons again pair up. The exact conditions under which such a scenario takes place depends on the details of the shell structure. But there are a handful of triplets of nuclei for which this scenario is realized. For such triplets the first-order beta decay of X into Y is energetically prohibited, but the second-order beta decay of X into Z is possible \cite{GoeppertMayer:1935qp}:
\begin{equation}
X \to  \> Z + 2 e^- + 2 \bar{\nu}_e  \;\; .
\end{equation}
Such double beta decays with two neutrinos in the final state, $2\nu \beta \beta$, have been observed in a number of nuclei. 

Already in 1937 it was pointed out that, if the neutrinos are Majorana fermions, the neutrino emitted by one of the nucleons can be absorbed by another one \cite{Racah:1937qq}. The resulting process,
\begin{equation}
X \to  \> Z + 2 e^-  \;\; ,
\end{equation}
is called neutrinoless double beta decay, $0\nu \beta \beta$. This process has not yet been  experimentally observed. However, if one starts with a very large number of parent nuclei and waits patiently for a handful of them to undergo $0\nu \beta\beta$ decay, this process may well prove to be an exception to the rule that, when relativistic, Majorana neutrinos behave just like Dirac ones \cite{refEVR}.

For the Majorana neutrino exchange, the leptonic part of the amplitude comes from the operator 
\begin{equation}
L_{\mu \nu} = \sum_i \bar{e} (x) \gamma_{\mu} (1-\gamma_5) U_{ei} \nu_i (x) \overline{\nu^c_i} (y) U_{ei} \gamma_{\nu} (1+\gamma_5) e^c(y) \;\; ,  
\end{equation}
where the sum is over the neutrino mass eigenstates. 
Contraction of the two neutrino fields in the above tensor yields the neutrino propagator 
\begin{equation}
\frac{\gamma_{\mu} q^{\mu} -m_i}{q^2 - m_i^2} \;\; .
\end{equation}
The $\gamma_{\mu}q^{\mu}$ term does not contribute to the traces, leaving the leptonic tensor proportional to the quantity 
\begin{equation}
m_{\beta \beta} \equiv \sum_i m_i U_{ei}^2 \;\; . 
\end{equation}

Calculation of the hadronic part of the amplitude, a nuclear matrix element, is significantly more complicated. Both $2 \nu \beta \beta$ and $0\nu \beta \beta$ decay modes involve virtual intermediate states of the nucleus Y. When two real neutrinos are emitted, the virtual momentum transfers are relatively small. The ground states of the even-even nuclei X and Z have spin-parity $0^+$. Hence to calculate the $2 \nu \beta \beta$ rate it is sufficient to include transitions through just a few low-lying $1^+$ states in the nucleus Y. In contrast, when the neutrinos remain virtual, as in the $0 \nu \beta \beta$ decay, virtual momentum transfers can reach to values of up to a few hundred MeV, necessitating inclusion of transfers through many intermediate states, including most of the particle-hole excitations in the nucleus Y. Hence the nuclear matrix elements in the 
$2 \nu \beta \beta$ and $0 \nu \beta \beta$ decay would be significantly different. For a comprehensive review of the nuclear matrix elements in double beta decay, we refer the reader to Ref.~\cite{Vogel:2012ja}. 

We can then write the half-life for the neutrinoless double beta decay where a Majorana neutrino is exchanged as 
\begin{equation}
\left[ T^{0 \nu \beta \beta}_{1/2} \left( 0^+_X \to 0^+_Z \right) \right]^{-1} = G_{0\nu} \>  |M_{0 \nu}|^2 \> |m_{\beta \beta}|^2  \;\; ,
\end{equation}
where $M_{0 \nu}$ is the appropriate nuclear matrix element and $G_{0 \nu}$ contains all the other easily calculable contributions, including phase space factors. The two-neutrino double beta decay rate can also be written in a similar form: 
\begin{equation}
\left[ T^{2 \nu \beta \beta}_{1/2} \left( 0^+_X \to 0^+_Z \right) \right]^{-1} = G_{2\nu} \>  |M_{2 \nu}|^2 \;\; .
\end{equation}

Note that there are considerable uncertainties coming from both nuclear physics and neutrino physics in the calculated neutrinoless double beta decay rates. The factor $m_{\beta \beta}$ depends on neutrino masses themselves, not their squares. This makes  $0 \nu \beta \beta$ decay 
exquisitely sensitive to the neutrino mass hierarchy and possible presence of additional mass eigenstates which mix into the electron neutrino. Also, because of the Majorana character of the neutrinos, $m_{\beta \beta}$ depends on the square of the mixing matrix element, not on its absolute value squared. As a result, the $0 \nu \beta \beta$ rate also depends on the phases in the mixing matrix, including two (or more if there are sterile states) additional phases that do not come into the neutrino oscillations. Those phases could interfere either constructively or destructively.

Since a neutrinoless double beta decay experiment can observe only the daughter nucleus and two electrons, the exchanged particle does not have to be a light neutrino. Lepton number violating interactions taking place at a scale $\Lambda$ above the electroweak scale could lead to exchange of heavier particles instead of the light neutrino. The contribution of such a heavy particle to the $0 \nu \beta \beta$ decay amplitude is roughly $\sim G_F^2 M_W^4/ \Lambda^5$ whereas the light neutrino exchange contributes a factor of $\sim G_F^2 m_{\beta \beta}/ k^2$, where $k$ is the exchanged virtual momentum, of the order of a few 100 MeV as we mentioned above. These two contributions become comparable at about $ \Lambda = 1 $ TeV. Calculations using effective field theory also come up with a similar scale \cite{Cirigliano:2004tc}. 

Experimental observation of $0 \nu \beta \beta$ decay, no matter what the underlying mechanism is, would imply that nature contains a Majorana neutrino mass, and that, therefore, neutrinos are Majorana fermions. This is because the decay implies that the lepton number violating amplitude converting two $d$ quarks into two $u$ quarks plus two electrons is non-zero. 
If this amplitude is not zero then each initial $d$ quark and final $u$ quark pair can be contracted to a $W$ boson. The two $W$ bosons can then combine with the two electrons in the final state. The resulting diagram is nothing but a contribution to the Majorana neutrino mass. 
This was pointed out a long time ago in Ref.~\cite{Schechter:1981bd}.

\subsection{An exotic exception}

A different kind of exception to the rule that ultra-relativistic Majorana and Dirac neutrinos behave indistinguishably can occur if there exists a heavy neutrino mass eigenstate $N$ satisfying 
$m_e \ll m_N \ll m_K$, where $m_e, \; m_N$, and $m_K$ are the electron, $N$, and kaon masses, respectively \cite{refK.4}. Since leptons mix, we would expect this $N$ to be a (small) component of $\nu_e$. Then the $N$ can be produced by the decay $K^+ \ra e^+ + N$, driven by the SM weak interaction. In this decay, because the kaon is spinless, the kaon-rest-frame helicities of the $e^+$ and $N$ must be of the same sign. Since $m_e \ll m_N \ll m_K$, the left-handed chiral projection operator in the SM weak interaction will almost always give the $e^+$ the usual right-handed helicity of a relativistic antilepton, forcing the $N$ to have right-handed helicity as well. If, for example,  $m_N = 50$ MeV, the $N$ will have right-handed helicity 99.99\% of the time. Now suppose this $N$ undergoes a charged-current or neutral-current SM weak interaction with some target. If neutrinos are Dirac particles, lepton number is conserved, so the $N$ from $K^+$ decay will be a neutrino, not an antineutrino. Given its right-handed helicity, its interaction will then be extremely suppressed by the left-handed chiral projection operator in the SM weak interaction. However, if neutrinos are Majorana particles, it can interact just like a right-handed Dirac antineutrino would, and there is no suppression.

\subsection{The special role non-relativistic neutrinos could play}   \label{sec3.5}

As we have seen, in almost all situations where a neutrino is relativistic, its helicity is a substitute for lepton number, so that its behavior will not reveal whether it is a Dirac or a Majorana particle. However, if the neutrino is {\em non-relativistic} instead, its behavior can depend quite a lot on whether it is a Dirac or a Majorana particle. As an illustration, let us consider the capture of the relic neutrinos from the Big Bang on tritium. At their current temperature, these neutrinos have $kT = 1.7 \times 10^{-4}$ eV. Given the known values of $\Delta m^2_{32}$ and $\Delta m^2_{21}$, if the mass ordering is normal, Mass$(\nu_{3}) \geq 5.0 \times 10^{-2} $ eV and Mass$(\nu_{2}) \geq 8.6 \times 10^{-3} $ eV. If the ordering is inverted, Mass$(\nu_{2} \;\mathrm{and}\;  \nu_{1}) \geq 5.0 \times 10^{-2} $ eV. 
Thus, for either ordering, two of the three known mass eigenstates are non-relativistic, and if the lightest member of the spectrum is not too light, all three of them are. Neglecting the small kinetic energy of one of the non-relativistic mass eigenstates, $\nu_i$, the capture of this mass eigenstate on a tritium nucleus via the reaction $\nu_i + \,^3\rm{H } \ra \,^3\rm{He} + e^-$ will yield a mono-energetic electron with an energy $E_e \cong (m_\mathrm{H} - m_\mathrm{He}) + m_{\nu_i}$, where $m_\mathrm{H}$, $m_\mathrm{He}$, and $m_{\nu_i}$ are the masses of the two participating nuclei and the neutrino, respectively. In contrast, the $\beta$ decay of a tritium nucleus yielding the lightest neutrino mass eigenstate $\nu_\mathrm{Li}$, $^{3}\mathrm{H} \rightarrow ^3\!\mathrm{He} + e^{-} + \overline{\nu_\mathrm{Li}}$, will yield an electron with energy $E_{e} \leq (m_\mathrm{H} - m_\mathrm{He}) - m_{\nu_\mathrm{Li}}$. To prove that relic neutrinos are being captured, an experiment must have sufficient energy resolution to establish that some of the electrons it observes have energies very slightly beyond the endpoint of the electron energy spectrum from $\beta$ decay. 

The relic neutrinos were highly relativistic when they were produced in the hot early universe, and the SM interactions that produced them yielded the same number of particles with negative helicity as with positive helicity. The number of particles produced with each helicity did not depend on whether neutrinos are Dirac or Majorana particles, since as we have seen Dirac and Majorana neutrinos behave identically when they are relativistic. Of course, if neutrinos are Dirac particles, then the relics created with negative helicity were (and still are) {\em neutrinos}, while those created with positive helicity were {\em antineutrinos}. After decoupling, the neutrinos free streamed, and as the universe expanded and cooled, many of them, and possibly all, became non-relativistic. Equality between the number of negative-helicity particles and positive-helicity ones was preserved during this evolution. 

For either Dirac neutrinos (not antineutrinos) or Majorana neutrinos, the amplitude for the capture of relic mass eigenstate $\nu_i$ on tritium obeys 
\beq
\mathrm{Amplitude}(\nu_i + ^3\!\mathrm{H }\ra \,^3\mathrm{He} + e^-) \propto \overline{u_e} \gamma^\lambda (1-\gamma_5) u_{\nu_i} J_\lambda^\mathrm{Nuclear} \;\;.
\label{eq22}
\eeq
Here, $u_e$ and $u_{\nu_i}$ are Dirac wave functions for the electron and the neutrino, respectively, and $J^\mathrm{Nuclear}_\lambda$ is a current describing the nuclear part of the process. The product $(1 -\gamma_5)u_{\nu_i}$ leads to a factor
\beq
1 - 2h_{\nu_i} \sqrt{(E_{\nu_i} - m_{\nu_i}) / (E_{\nu_i} + m_{\nu_i})} \equiv F(h_{\nu_i},E_{\nu_i})\label{eq23}
\eeq                   
in the amplitude, where $E_{\nu_i}$ is the energy of the neutrino, and $h_{\nu_i}$ is its helicity. Now, suppose the relic neutrinos were still highly relativistic, with $E_ {\nu_i} \gg {m_ {\nu_i}}$, in the rest frame of the tritium today. Then, if neutrinos are Majorana particles, capture of the half of the relic population with $h_{\nu_i}= +1/2$ would be extremely suppressed by $F(h_{\nu_i}, E_{\nu_i})$. (It can be shown that the details of $J^\mathrm{Nuclear}_\lambda$ do not affect this argument. However, if we view the Lorentz-invariant amplitude of \Eq{22} from the rest frame of the neutrino, in which $F(h_{ \nu_i}, E_{\nu_i}) = 1$, but in which the target nucleus is moving at high speed, the details of $J^\mathrm{Nuclear}_\lambda$ are all important, and lead to the same suppression of capture that we find when viewing the amplitude from the rest frame of the target \cite{refK.6}.)  If neutrinos are Dirac particles, the half of the relic population with $h_{\nu_i}= +1/2$ cannot be captured by tritium to make an electron because the positive-helicity relics are antineutrinos, and lepton number is conserved. Capture of the half of the relic population with $h_{\nu_i}= - 1/2$ would be described by the amplitude of \Eq{22} in either the Dirac or Majorana case, and would not be suppressed. Thus, the event rate would have no visible dependence on whether neutrinos are Dirac or Majorana particles. 

In reality, as we have discussed, many, and perhaps all, of the relic neutrinos have become non-relativistic in the tritium/laboratory rest frame. For the non-relativistic relics, $F(h_{\nu_i}, E_{\nu_i}) \cong 1$, causing little suppression and depending very little on the neutrino helicity. If neutrinos are Majorana particles, the amplitude for capture of a neutrino with either positive or negative helicity is given by \Eq{22}, and with $F(h_{\nu_i}, E_{\nu_i}) \cong {1}$ independent of the helicity, the relic populations with positive and negative helicity will contribute equally to the capture rate. If neutrinos are Dirac particles, the amplitude for capture of a neutrino with negative helicity is again given by \Eq{22}, and is the same as in the Majorana case, but because of lepton number conservation, the amplitude for capture of a neutrino with positive helicity, which is an antineutrino, is zero. Thus, the total capture rate is twice as large in the Majorana case as in the Dirac case \cite{refK.6a}. 

While this dependence of the capture rate on whether neutrinos are Dirac or Majorana particles is very substantial, it must be acknowledged that actually using relic capture on tritium to determine whether neutrinos are of Dirac or Majorana character faces major challenges. First of all, the observation of this capture is very difficult, and has not yet been accomplished \cite{refK.5}. Secondly, the capture rate obviously depends not only on the cross section for the process, but also on the local density of relic neutrinos. Owing to gravitational clustering, this local density could be very different from the average density in the universe as a whole, and is much less precisely predicted than the latter  \cite{refK.7}. Thirdly, if the lightest neutrino mass eigenstate is light enough to be relativistic today, finite experimental energy resolution could well make it impossible to tell that an electron from its capture is not one from tritium $\beta$ decay. Thus, its capture would not be counted. Now, $|U_{e1}|^{2} \cong 2/3$, so if the mass ordering is normal, so that the lightest mass eigenstate is $\nu_1$, two-thirds of the captures would not be counted.

\subsection{Angular distributions in decays}

The search for non-relativistic neutrinos whose behavior might be revealing leads us to consider neutrino decays, since of course a neutrino undergoing decay is totally non-relativistic in its rest frame. Let us first consider the decay of a Majorana neutrino $\nu_{2}^{(M)}$ into another Majorana neutrino $\nu_1^{(M)}$ and a photon:
\begin{equation}
\nu_2^{(M)} \to \nu_1^{(M)} + \gamma \;\; .
\end{equation}
Angular momentum conservation implies that the amplitude of such a process in the helicity formalism is given by
\begin{equation}
\label{helicitycond}
D^{j*}_{m ,\lambda} (\phi, \theta, - \phi) A_{\lambda_1, \lambda_{\gamma}} \;\; .
\end{equation}
Here, $D$ is the Wigner rotation function, $j,m$ are the spin and third component of the spin along the z-axis for the decaying particle $\nu_2^{(M)}$, $\lambda_1$ and 
$\lambda_{\gamma}$ are the helicities of the decay products and $\lambda = \lambda_{\gamma} - \lambda_1$. With no loss of generality we assume that $\nu_2^{(M)}$ is polarized in the +z direction and evaluate the decay amplitude in its rest frame (see {\bf Figure \ref{majorana_decay}}). In this configuration, from  angular momentum conservation it follows that 
\begin{equation}
\label{angmomcons}
|\lambda_{\gamma} - \lambda_1| \le j = 1/2 \;\;  .
\end{equation}
\begin{figure}[h]
\includegraphics[width=4in]{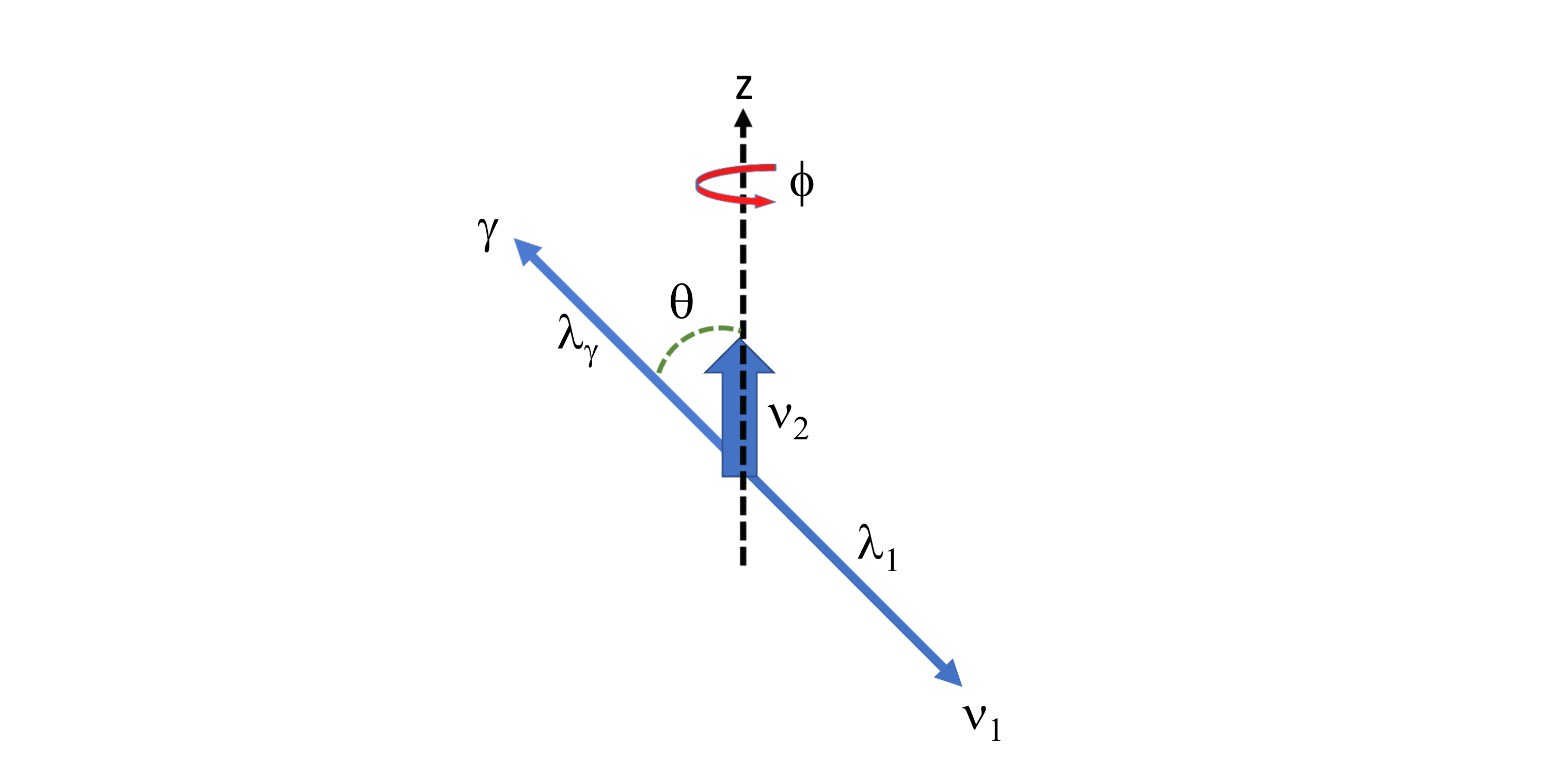}
\caption{Decay configuration of the Majorana fermion.}
\label{majorana_decay}
\end{figure}
There are two possible helicities for the photon, both of which contribute for Majorana fermions. For the case of $\lambda_{\gamma} = +1$ , Eq.~(\ref{angmomcons}) implies $\lambda_1= +1/2$, hence $\lambda = +1/2$. For $\lambda_{\gamma} = -1$ , one has $\lambda_1= -1/2$, hence $\lambda = -1/2$. The tree level amplitudes are then
\begin{equation}
\label{helicity+1}
\langle \gamma (\mathbf{p}, +1) \> \nu_1 (\mathbf{-p}, +1/2) | H_{\rm EM} | \nu_2(0, +1/2) \rangle = d_{+1/2 , +1/2}^{1/2}  A_{+1, +1/2}
\end{equation}
and 
\begin{equation}
\label{helicity-1}
\langle \gamma (\mathbf{p}, -1) \> \nu_1 (\mathbf{-p}, -1/2) | H_{\rm EM} | \nu_2(0, +1/2) \rangle = d_{+1/2 , -1/2}^{1/2}  A_{-1, -1/2}
\end{equation}
up to phases not explicitly shown. Here, $H_{\rm EM}$ is the effective Hamiltonian for the decay, and $d^j_{m\lambda}$ is the reduced Wigner rotation function. Imposing the condition of invariance under CPT transformation \cite{Bell:2008fm}, described by the operator $\zeta$,  one can write 
\begin{eqnarray}
\langle \gamma (\mathbf{p}, \lambda_{\gamma}) \> \nu_1 (\mathbf{-p}, \lambda_1) | H_{\rm EM} | \nu_2(0, +1/2) \rangle 
&=& \langle \zeta H_{\rm EM} \zeta^{-1} \zeta [\nu_2(0, +1/2)] | \zeta [  \gamma (\mathbf{p}, \lambda_{\gamma}) \> \nu_1 (\mathbf{-p}, \lambda_1)]    \rangle \nonumber \\
&=& \langle \gamma (\mathbf{p}, -\lambda_{\gamma}) \> \nu_1 (\mathbf{-p}, -\lambda_1) | H_{\rm EM} | \nu_2(0, -1/2) \rangle^*
\label{CPTinv}
\end{eqnarray}
up to phases not explicitly shown. Substituting the values $\lambda_{\gamma} = +1$,  
$\lambda_1 = +1$  in the first and the third entries in Eq.~(\ref{CPTinv})
and using Eq.~(\ref{helicitycond}) we get 
\begin{equation}
\label{A-=A+}
d_{+1/2,+1/2}^{1/2} A_{+1,+1/2} = d_{-1/2,-1/2}^{1/2}  A^*_{-1, -1/2} \Rightarrow  A_{+1,+1/2} = A^*_{-1, -1/2}\equiv A
\end{equation}
up to phases not shown. The first order decay rate into photons with helicity $\lambda_{\gamma} = +1$ can be read from Eq.~(\ref{helicity+1}) to be 
\begin{equation}
\frac{d\Gamma_+}{d \cos \theta} = \left( d_{+1/2 , +1/2}^{1/2} \right)^2  |A_{+1, +1/2}|^2 = \cos^2 \frac{\theta}{2} |A_{+1, +1/2}|^2 
\end{equation}
Similarly the decay rate into photons with helicity $\lambda_{\gamma} = -1$ from Eq.~(\ref{helicity-1}) is 
\begin{equation}
\frac{d\Gamma_-}{d \cos \theta} = \left( d_{+1/2 , -1/2}^{1/2} \right)^2  |A_{-1, -1/2}|^2 = \sin^2 \frac{\theta}{2} |A_{-1, -1/2}|^2 
\end{equation}
Summing over final helicities we find the total decay rate for a spin-up Majorana fermion to be 
\begin{equation}
\frac{d\Gamma}{d \cos \theta} = \cos^2 \frac{\theta}{2} |A_{+1, +1/2}|^2 + \sin^2 \frac{\theta}{2} |A_{-1, -1/2}|^2 
\end{equation}
which, using Eq.~(\ref{A-=A+}), takes the form 
\begin{equation}
\frac{d\Gamma}{d \cos \theta} =|A|^2 \;\; , 
\end{equation}
i.e., the distribution of photons is isotropic.
Note that this isotropy is a consequence of angular momentum conservation and CPT invariance alone. \cite{refK.8} It does not depend on any further details of the interactions involved in the decay. 

In contrast to the Majorana case, one finds by explicit calculation that if neutrinos are Dirac particles, the radiative decay $\nu_{2}^{(D)} \rightarrow \nu_{1}^{(D)} + \gamma$ of polarized neutrinos $\nu_2^{(D)}$ will in general yield a non-isotropic distribution of photons if the decay is driven by both magnetic and electric transition dipole moments. Thus, the angular distribution of photons from radiative neutrino decay can in principle be used to determine whether neutrinos are Dirac or Majorana particles. \cite{refK.9} Should it be less challenging to measure the polarization of these photons than their angular distribution, the polarization can also be used for this purpose. If neutrinos are Majorana particles, their helicity will be $\cos\theta$, which, given their isotropic angular distribution, will result in an angle-integrated helicity of zero \cite{refK.9}. If they are Dirac particles, their angle-integrated helicity will in general not be zero. 

What if there exists an as-yet-undiscovered neutrino $N$ that is much heavier than the three known ones? \cite{refK.10} Then there will be new and potentially quite revealing decay modes. Among these are decays of the form $N\rightarrow {\nu} + X$, where $\nu$ is one of the three known light neutrino mass eigenstates, and $X = \overline{X}$ is a particle that is identical to its antiparticle. Depending on the mass of $N$, $X$ can be, for example, a $\gamma, \pi^0, \rho^0, Z^0$, or $H^0$. What these modes could teach us is analyzed in Ref.~\cite{AnBaBo}. For each of them, if the neutrinos, including $N$, are Majorana particles, the decay rate is twice as large as it is if the neutrinos are Dirac particles \cite{refK.11}. However, the decay rate for any given mode also depends on unknown mixing angles, so a measurement of the rate may not reveal whether neutrinos are of Dirac or Majorana character. Therefore, it is quite interesting that the angular distribution of the outgoing particles in these decays is another feature that depends on whether neutrinos are Dirac or Majorana particles. Furthermore, except when $X = \gamma$, this dependence does not involve elusive parameters.

Let us assume that the reaction producing the heavy neutrino $N$ in some experiment leaves it fully polarized with its spin in the $+z$ direction in its rest frame. Let us also assume, first, that $N$ is a Majorana particle $N^{(M)}$ and $\nu$ is a Majorana particle $\nu^{(M)}$ (and $X = \overline{X}$ is a self-conjugate particle as well). Then, through a generalization of the analysis given above for $\nu_2^{(M)}\rightarrow \nu_1^{(M)} +\gamma$, we can show that, purely as a result of angular momentum conservation and CPT invariance, the angular distribution of $X$ particles from the decay $N^{(M)}\rightarrow \nu^{(M)} +X$ will be isotropic in the $N^{(M)}$ rest frame. Next, let us assume instead that $N$ is a Dirac particle $N^{(D)}$, $\nu$ is a Dirac particle $\nu^{(D)}$, and X is the same self-conjugate particle as before. Then the angular distribution of $X$ particles from $N^{(D)}\rightarrow \nu^{(D)} +X$ in the $N^{(D)}$ rest frame will be 
\beq
\frac{d\Gamma}{d(\cos \theta)} = \Gamma_0 (1 + \alpha \cos \theta) \;\;,
\label{eqK.24}
\eeq
where  $\Gamma_0$ is a normalization constant, and $\alpha$ is given for each $X$ considered in Ref.~\cite{AnBaBo} in Table \ref{tabK1}. 
\begin{table}[h]
\caption{The coefficient $\alpha$ in the angular distribution $(1+ \alpha\cos\theta)$ of the particle $X$ from the decay $N^{(D)}\rightarrow \nu^{(D)} +X$ of a heavy Dirac neutrino $N^{(D)}$ fully polarized with its spin in the $+z$ direction. The quantities $\mu$ and $d$ are, respectively, the magnetic and electric transition dipole moments that drive $N^{(D)}\rightarrow \nu^{(D)} +X$, and $m_N,\, m_\rho$, and $m_Z$ are, respectively, the masses of $N^{(D)}, \rho^0$, and $Z^0$.}
\label{tabK1}
\Large
\begin{center}
\scalebox{1.6}{
\begin{tabular}{|c||c|c|c|c|c|}
\hline
X		& $\gamma$		& $\pi^0$	& $\rho^0$		& $Z^0$		& $H^0$	\\
\hline  
$\alpha$	&  $\frac{2\Im m (\mu d^*)}{|\mu|^2+|d|^2}$		& $1$	
		&  {\rule[-4mm]{0mm}{9mm} $\frac{m^2_N-2m_\rho^2}{m^2_N+2m_\rho^2} $ }
		&  $\frac{m^2_N-2m_Z^2}{m^2_N+2m_Z^2}$	& $1$	\\
\hline
\end{tabular}}  
\end{center}  
\end{table}
In the calculations summarized by this table, it has been assumed that when $X = \pi^0$ or $\rho^0$, the decay is dominated by a virtual $Z^0$ that is emitted by the lepton line and becomes the $X$, and that the coupling of the $H^0$ to the lepton line is a Yukawa coupling. In the corresponding antineutrino decays, $\overline{N^{(D)}} \rightarrow \overline{ \nu^{(D)}} +X$, the angular distribution is the same except for a reversal of the sign of $\alpha$.

As Table \ref{tabK1} shows, except under very special circumstances, such as $m_N^2 = 2m_\rho^2$, the angular distribution in the Dirac case is not isotropic. When $X$ is a $\rho^0$ or a $Z^0$, the only presently-unknown parameter in this distribution is the mass of $N$, which would likely be measured once $N$ is discovered. When $X$ is a $\pi^0$ or an $H^0$, the angular distribution does not depend on any parameters at all. Thus, the study of angular distributions in the decays of a heavy neutrino could tell us whether neutrinos are Dirac or Majorana particles. 

Imagine, finally, that a heavy neutrino $N$ is created together with an $e^+$ in the decay of some particle that is not a lepton. If it is found that this same $N$ can undergo the decay $N\rightarrow e^+ + \pi^-$, then lepton-number conservation is obviously violated and this $N$ must be a Majorana particle. Needless to say, chains of events such as this are very well worth searching for.

\section{ELECTROMAGNETIC STRUCTURE OF NEUTRINOS}

Having considered whether neutrinos are of Majorana or Dirac character, we now turn to their electromagnetic structure. As we shall see, this structure is not independent of their Majorana vs.\ Dirac nature. 

The most general matrix element of the electromagnetic current $J_\mu^{EM}$ between neutrino mass eigenstates $\nu_i$ and $\nu_j$ is given by
\begin {eqnarray}  
\left\langle \nu_j(p_j) | J^{EM}_\mu | \nu_i(p_i)  \right\rangle   =   \bar{u}_j(p_j) \;\;  \times \hspace{3.0in}  &  & \nonumber   \\
	 \times  \left\{ \left( \gamma_\mu - q_\mu \frac{\gamma_\nu q^\nu}{q^2} \right)
	 [f_Q^{ji} (q^2) + f_A^{ji}(q^2) q^2\gamma_5]     
	-  i\sigma_{\mu\nu} q^\nu [f_M^{ji}(q^2) + if_E^{ji}(q^2)\gamma_5] \right\} u_i(p_i) \;  .  
\label{eqK.25}
\end{eqnarray}
Here, $q = p_i  - p_j$ is the momentum transfer, and the various factors $f$ are Hermitean matrices of form factors. In particular, the matrices $f_Q, f_M, f_E$, and $f_A$ contain the charge, magnetic dipole, electric dipole, and anapole form factors, respectively. For the coupling to a real photon $(q^2 = 0), \; f_M$ and $f_E$ reduce to transition (if $j \neq i$) or intrinsic (if $j = i$) magnetic and electric dipole moments, respectively. \cite{refK.13, Giunti:2014ixa}

Excellent comprehensive reviews of neutrino electromagnetic structure are available in the literature \cite{Giunti:2014ixa,Giunti:2015gga}.
Hence in this section we will limit our discussion to those aspects pertinent to neutrino magnetic moments.

In the minimally extended  Standard Model, which includes neutrino masses and mixing, the neutrino magnetic moment is very small \cite{Lee:1977tib,Fujikawa:1980yx}. For Dirac neutrinos, the magnetic moment matrix in the mass basis is given by 
\begin{equation}
\mu_{ij} = - \frac{1}{\sqrt{2}} \frac{eG_F}{8 \pi^2} \left( m_i + m_j \right) \sum_{\ell} U_{\ell i} U^*_{\ell j} f(r_{\ell}) \;\; ,
\end{equation}
where
\begin{equation}
f(r_{\ell}) \sim - \frac{3}{2} + \frac{3}{4} r_{\ell} + \cdots , \>\>\>\>\> r_{\ell} = \left( \frac{m_{\ell}}{M_W} \right)^2  \,\, ,
\end{equation}
and $m_\ell$ is the mass of charged lepton $\ell$. Since three neutrino mixing angles and upper bounds on neutrino mass are known, one can calculate this SM prediction  as a function 
of the unknown neutrino masses $m_i$ and $m_j$ and demonstrate that it is well below experimental reach \cite{Balantekin:2013sda, refF}.

\subsection{Neutrino-electron scattering}

The differential scattering
cross section for $\nu_e$s and $\overline{\nu}_e$s  
on electrons is given by (see e.g. Ref.~\cite{Vogel:1989iv})
\begin{eqnarray}
\frac{d\sigma}{dT} &=& \frac{G_F^2m_e}{2\pi} 
\left[(g_V + g_A)^2 + (g_V - g_A)^2 \left(1-\frac{T}{E_\nu}\right)^2 + 
(g_A^2-g_V^2) \frac{m_eT}{E_\nu^2}\right] \nonumber \\
&+& \frac{\pi \alpha^2 \mu_{\nu}^2}{m_e^2} \left[ \frac{1}{T} - 
\frac{1}{E_{\nu}} \right], 
\label{d1}
\end{eqnarray}
where T is the electron recoil kinetic energy, $g_V = 2 \sin^2 \theta_W 
+ 1/2$, $g_A = +1/2 (-1/2)$ for $\nu_e$ ($\overline{\nu}_e$), and the 
neutrino magnetic moment is expressed in units of Bohr magneton, $\mu_B$. The first line 
in Eq. (\ref{d1}) is the weak and the second line is the magnetic moment contribution\footnote{The contribution of the interference 
of the weak and magnetic amplitudes to the cross section is proportional to the neutrino mass and can be ignored for ultrarelativistic 
neutrinos.}. Experiments searching for the  neutrino dipole moments utilize the fact that the magnetic moment cross section exceeds the weak cross-section for recoil energies
\begin{equation}
\frac{T}{m_e} < \frac{\pi^2 \alpha^2}{(G_F m_e^2)^2} \mu_{\nu}^2 \;\; .
\label{d2}
\end{equation}  
That is, the lower the smallest measurable recoil energy is, the smaller the
values of the magnetic moment that can be probed. 
A reactor experiment measuring the antineutrino magnetic moment by detecting the electron recoil is an inclusive one, i.e.\ it sums over all the neutrino final states.  It should be noted that, because neutrinos oscillate between their source and the detector, the value of the $\mu_{\nu}$ of 
Eq.~\ref{d1} measured at a distance $L$ from the neutrino source is an effective value: 
\begin{equation}
\mu_{\rm eff}^2 = \sum_i \left|  \sum_i U_{ej} \mu_{ij} \exp (-i E_j L)\right|^2 \;\; ,
\label{reactormunu}
\end{equation}
where $i,j$ are mass indices, $ \mu_{ij}$ is the dipole moment matrix in the mass basis, and $U_{ej}$ are elements of the neutrino mixing matrix. Currently the best reactor neutrino limit is given by the GEMMA spectrometer at  Kalinin Nuclear Power Plant to be 
$\mu_{\nu} < 2.9 \times 10^{-11} \mu_B$ \cite{Beda:2013mta}. 

There are small radiative loop corrections to the tree diagrams calculated in Eq.~\ref{d1}
\cite{Marciano:2003eq}. A comparison of the tree-level weak, magnetic and radiative corrections to the tree-level weak contribution to the cross section is given in {\bf Figure \ref{fignicole}}, taken from Ref.~\cite{Balantekin:2013sda}. 
\begin{figure}[h]
\includegraphics[width=4in]{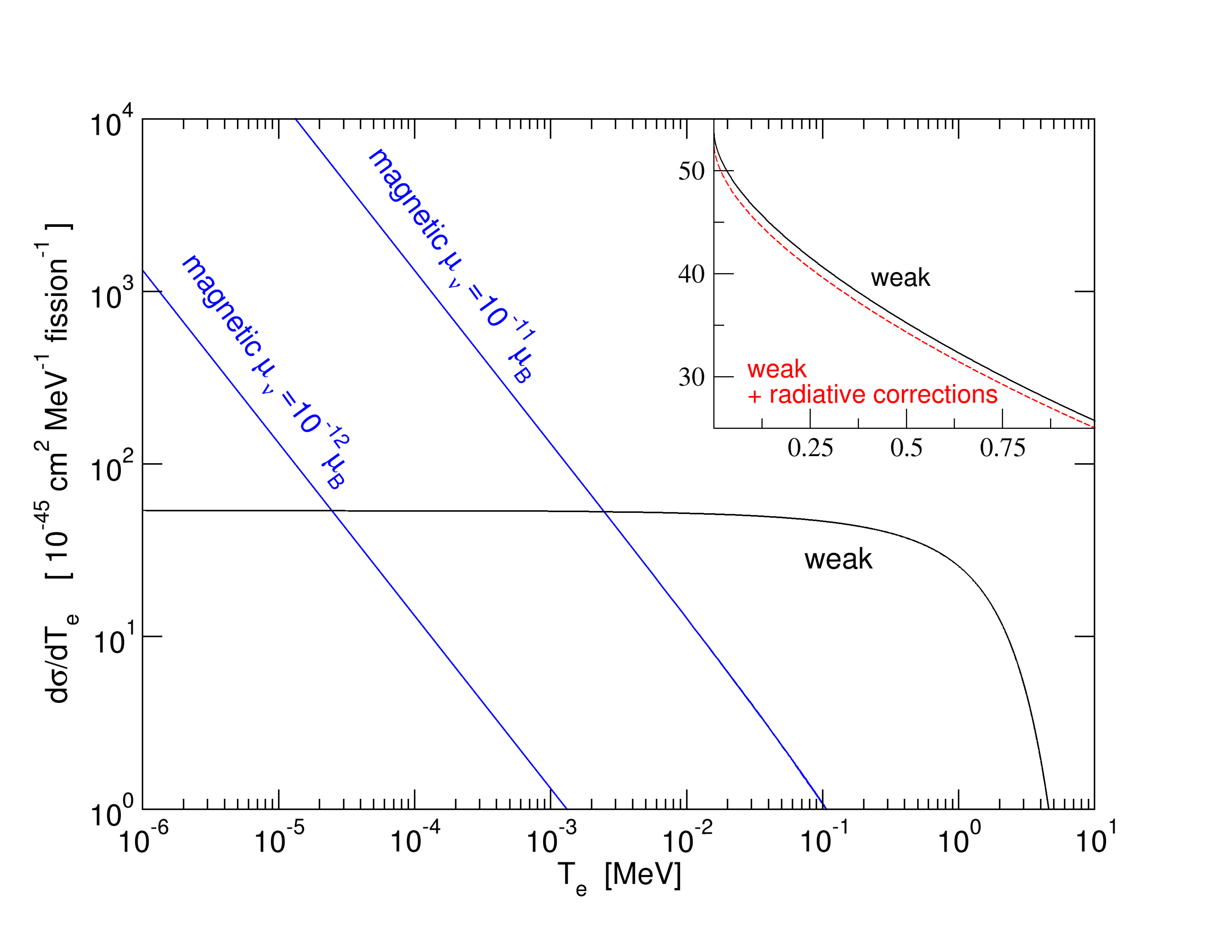}
\caption{A comparison of the tree-level weak, magnetic and radiative corrections to the tree-level weak contribution to the differential cross section is for electron antineutrino-electron scattering (taken from Ref.~\cite{Balantekin:2013sda}). 
The axis labels of the insert are the same as the axis labels of the larger figure.}
\label{fignicole}
\end{figure}

\subsection{Effects of magnetic moments in neutrino propagation}

The precession of neutrino spins in magnetic fields that is induced by magnetic moments has been studied in Ref.~\cite{Fujikawa:1980yx}. 
The rotation
\begin{equation}
\nu_{eL} \to \nu_{eR} 
\end{equation}
produces a right-handed neutrino when magnetic fields transverse to the direction of neutrino propagation are present \cite{Cisneros:1970nq,Okun:1986na}. 
It was subsequently realized that matter effects would break the vacuum degeneracy of the
$\nu_{eL}$ and $\nu_{eR}$ states, suppressing the spin precession shown above.  However, it was 
pointed out \cite{Lim:1987tk,Akhmedov:1988uk} that this
difficulty was naturally circumvented for the process 
\begin{equation}
\nu_{eL} \to \nu_{\mu R} 
\end{equation}
as different matter interactions of the $\nu_e$ and $\nu_\mu$ can
compensate for the vacuum $\nu_e - \nu_\mu$ mass difference, producing
a crossing similar to the usual MSW mechanism.  Such spin-flavor
precession can then occur at full strength due to off-diagonal magnetic moments with flavor indices. Note that only the product of the transverse magnetic field and 
neutrino magnetic moments appears in the equations describing spin-flavor precession.  
Spin-flavor precession in the Sun was studied in detail \cite{Balantekin:1990jg}, 
motivated by the Homestake solar neutrino data that suggested an anti-correlation between solar neutrino capture rate and the number of sunspots, a proxy of the magnetic activity in the Sun. Although this correlation was weakened by further observations, it was realized that spin-flavor precession would produce solar antineutrinos if the neutrinos are of Majorana type 
\cite{Raghavan:1991em}. Solar neutrino experiments searching for such antineutrinos report null results \cite{Liu:2004ny,Borexino:2017fbd}. 
The physics of the Sun does not seem to be affected by neutrino magnetic moments \cite{Balantekin:2004tk}, but can be used to place limits on the effective neutrino magnetic moments, yielding $\mu_{\rm eff} < 2.8 \times 10^{-11} \mu_B$  \cite{Borexino:2017fbd,Canas:2015yoa}. In contrast, neutrino magnetic moments may play a role in the evolution of massive stars \cite{Heger:2008er}. 

A good fraction of the heavier nuclei were formed in the rapid neutron 
capture (r-process) nucleosynthesis scenario. One expects the astrophysical sites of the r-process 
to be associated with explosive phenomena, since 
a large number of interactions are required to take place during a 
rather short time interval. Leading candidates include neutron star mergers and core-collapse supernovae. Several orders of magnitude greater enhancement of r-process element abundances was observed in an ultra-faint dwarf (i.e.\ very old) galaxy than has been seen in other such galaxies, implying that a single rare event produced the r-process material \cite{ji}, an argument in favor of neutron star mergers. 
A signature of nucleosynthesis in the neutron-star mergers would be the electromagnetic transients from the decay of radioactive isotopes they would produce  \cite{Martin:2015hxa}. 
The LIGO and Virgo collaborations reported observation of gravitational waves from a binary neutron star merger \cite{ligovirgo}. Multi-messenger observations of this 
binary neutron star merger established the presence of an electromagnetic counterpart \cite{multimessenger}, further supporting neutron star mergers as a site, but not ruling out core-collapse supernovae as another possible secondary site. 
The salient point for our subject is that both of these sites contain copious amounts of neutrinos. 
A key quantity for determining the r-process yields is the neutron-to-seed nucleus ratio, 
which, in turn, is determined by the neutron-to-proton ratio.  
The neutrino-induced processes such as  
$\nu_e + n \rightarrow p + e^-$ could significantly alter the
neutron-to-proton ratio. During the epoch of alpha-particle formation almost 
all the protons and an equal amount of neutrons combine into alpha particles 
which have a large binding energy. This ``alpha effect'' reduces the number of 
free neutrons participating in the r-process \cite{Fuller:1995ih,Meyer:1998sn}. 
Naively, one may assume that the presence of a neutrino magnetic moment would reduce the electron neutrino flux, resulting in a possible  mechanism to suppress alpha particle formation. 
This expectation is not realized since a non-zero magnetic moment suppresses the electron neutrino {\it and} antineutrino fluxes at the same time \cite{Balantekin:2007xq}. 
Large values of neutrino magnetic moments {\it would increase} the electron fraction and thus amplify the $\alpha$-effect.

Resonant neutrino spin-flavor precession in supernovae and its impact on nucleosynthesis has also been studied \cite{Nunokawa:1996gp}. 
Perhaps a more dominant effect in both neutron star mergers and core-collapse supernovae is 
 collective oscillations of neutrinos. These are emergent nonlinear flavor evolution phenomena instigated by neutrino-neutrino interactions in astrophysical environments with sufficiently high neutrino densities. 
There may be non-negligible effects of transition magnetic moments on three-flavor collective oscillations of Majorana neutrinos in core-collapse supernovae \cite{deGouvea:2013zp,deGouvea:2012hg}. Furthermore, the effects of neutrino dipole moments in collective oscillations 
are intertwined with the CP-violating phases of the neutrino mixing matrix \cite{Pehlivan:2014zua}.

\subsection{Other astrophysical and cosmological consequences}  

Studies of the red giant cooling process of plasmon decay into neutrinos,
\begin{equation}
\gamma^* \to \nu_i \bar \nu_j \;\; ,
\end{equation}
imposes limits on the neutrino dipole moments. A large enough neutrino magnetic moment would imply an enhanced plasmon decay rate. Since neutrinos freely escape the star, large neutrino dipole moments cool the red giant star faster, delaying helium ignition. 
The most recent such limit is  $|\mu_{\nu}| < 4.5 \times 10^{-12} \mu_B$ \cite{Viaux:2013lha}. 

Limits on the magnetic moments of Dirac neutrinos were given in the 1980's using cosmological arguments. Since magnetic moments contribute to  neutrino electron scattering and electron-positron annihilation into  neutrino pairs, large values of the magnetic moments would populate wrong-helicity counterparts, leading to an increase in N$_{\rm eff}$. These arguments limit the dipole moments of  Dirac neutrinos to be $\mu_{\nu} < 10^{-10} \mu_B$ \cite{Morgan:1981zy,Fukugita:1987uy}. However, such a consideration is restricted to Dirac neutrinos since Majorana neutrinos do not have additional neutrino states that can get populated by dipole moment-induced transitions. 

It is still possible to explore the impact of Majorana neutrino magnetic moments in the Early Universe.  Since the energy dependence of the weak and magnetic components of electron-neutrino scattering (cf.\ Eq.~\ref{d1}) are very different, they can have significantly different contributions to the reaction rates in the Early Universe. Hence, a sufficiently large magnetic moment can keep the Majorana neutrinos in thermal and chemical equilibrium below the standard ( $\sim 1$ MeV) weak decoupling temperature regime and into the Big Bang Nucleosynthesis (BBN) epoch. The production of light elements in the BBN epoch is very sensitive to the weak decoupling temperature, since the neutron-to-proton ratio is exponentially dependent on it. This high sensitivity can be exploited to obtain a limit on the effective neutrino magnetic moment through constraints on the observed primordial abundances, such as those of helium and deuterium. It follows that light element abundances and other cosmological parameters are sensitive to magnetic couplings of Majorana neutrinos on the order of $10^{-10}$ $\mu_B$ \cite{Vassh:2015yza}.

 \subsection{Neutrino decay in astrophysics}

So far, no evidence of neutrino decay has been observed in terrestrial experiments. However,
an unidentified emission line was seen in the X-ray spectrum of galaxy clusters: a monochromatic, 3.5 keV line in the X-ray spectrum that could be interpreted as a signal emerging from a decaying 7 keV sterile neutrino that mixes with active ones \cite{Bulbul:2014sua,Boyarsky:2014jta}. Such a neutrino can be resonantly produced in the Early Universe and constitute dark matter \cite{Abazajian:2017tcc}. 
If the sterile neutrino interpretation is indeed correct, the observed X-ray line would imply the presence of new entries in the the neutrino dipole moment and mixing matrices.  Sterile neutrino dark matter is expected to be in the range suggested by these observations \cite{Dolgov:2000ew} and may have been produced in the Early Universe \cite{Abazajian:2001nj}. Several planned missions dedicated to the search for X-ray lines from dark matter should elucidate the sterile neutrino decay interpretation of the 3.5 keV line.

\subsection{Magnetic moments of Dirac and Majorana neutrinos}

Neutrino mass and neutrino magnetic moment are not completely independent of one another. For example, one can write down a generic expression for the magnetic moment as
\begin{equation}
\mu_{\nu} \sim \frac{e {\cal G}}{\Lambda} \;\; ,
\end{equation}
where $\Lambda$ is the energy scale of the physics beyond the SM generating the magnetic moment at low energies, e is the charge on the electron, and ${\cal G}$ represents calculations of the appropriate diagrams connected to a photon. If this external photon is removed the same set of diagrams  
contribute a neutrino mass of the order 
\begin{equation}
\delta m_{\nu} \sim {\cal G} \Lambda  \;\; ,
\end{equation}
implying the relationship 
\begin{equation}
\label{deltameq}
\delta m_{\nu} \sim  \left( \frac{\mu_{\nu}}{\mu_B} \right) \frac{\Lambda^2}{2 m_e} \;\; .
\end{equation}
However, such a relationship can be circumvented using symmetry arguments, for example imposing a new symmetry that would force the neutrino mass to vanish \cite{Voloshin:1987qy}. 

A more robust connection can be obtained using effective field theory techniques. At lower energies, integrating out the physics above the scale $\Lambda$  one can write an effective Lagrangian consisting of local operators written in terms of the SM fields: 
\begin{equation}
{\cal L}_{\rm eff} = {\cal L}_{\rm SM} + \sum_{n=4}^{\cal N} \frac{1}{\Lambda^{n-4}} \sum_{j_n} C_{j_n}^{(n)}(\upsilon) {\cal O}_{j_n}^{(n)} (\upsilon)  \;\; ,
\end{equation}
where $n$ is the operator dimension, ${\cal N}$ specifies the number of the terms kept, $j_n$ labels all the independent operators of  dimension $n$, and $\upsilon$ is the renormalization scale used. As described earlier, to obtain a mass term for Dirac neutrinos one introduces a SM singlet field $\nu_R$ and writes a mass term of dimension four in a similar way to charged leptons. Using this additional SM field one can write three independent dimension six operators:
\begin{eqnarray}
{\cal O}_1^{(6)} &=& g \bar{L}\tau^a \bar{\epsilon} H^*\sigma^{\mu \nu} \nu_R \left(\partial_{\mu} W^a_{\nu} - \partial_{\nu} W^a_{\mu} - g \epsilon_{abc} W^b_{\mu} W^c_{\nu} \right) \nonumber \\
{\cal O}_2^{(6)} &=& g' \bar{L} \bar{\epsilon} H^*\sigma^{\mu \nu} \nu_R \left(\partial_{\mu} B_{\nu} - \partial_{\nu} B_{\mu}  \right) \nonumber \\
{\cal O}_3^{(6)} &=& \bar{L} \bar{\epsilon} H^* \nu_R \left( H^{\dagger} H \right)
\end{eqnarray}
where $L$ is the SM left-handed lepton isodoublet, $W$ and $B$ are the SU(2)$_L$ and U(1)$_Y$ gauge fields of the SM, and $\bar{\epsilon} = - i \tau_2$. Noting that $g= e/\sin \theta_W$ and $g' = e/\cos \theta_W$ one observes that, after the spontaneous symmetry breaking, the operators 
${\cal O}_1^{(6)}$ and ${\cal O}_2^{(6)}$ would generate a contribution to the magnetic dipole moment and ${\cal O}_3^{(6)}$ would generate a contribution to the neutrino mass. The appropriate renormalization group analysis was carried out in Ref.~\cite{Bell:2005kz}. Neglecting 
possible fine-tunings of the coefficients $C_j^{(6)}$, they found that a magnetic moment will rather generically induce a radiative correction to the Dirac neutrino mass of the order of
\begin{equation}
\delta m_{\nu} \sim \left( \frac{\mu_{\nu}}{10^{-15} \mu_B} \right) [\Lambda ( {\rm TeV} ) ]^2 {\rm eV} \;\; . 
\end{equation}
This bound was derived for a single flavor. For a hierarchical neutrino mass spectrum, it would be even more stringent. 

For Majorana neutrinos the analysis needs to be quite different. First of all one does not introduce a new SM field $\nu_R$. Instead, the Majorana mass term is given as a unique dimension five operator. Neutrino magnetic moments and corrections to the neutrino mass then come from dimension seven operators. A magnetic moment is generated by the operators 
\begin{eqnarray}
{\cal O}_1^{(7)} &=& g ( \overline{L^c} \bar{\epsilon} H ) \sigma^{\mu \nu} ( H^T
\bar{\epsilon} \tau_a L )  \left(\partial_{\mu} W^a_{\nu} - \partial_{\nu} W^a_{\mu} - g \epsilon_{abc} W^b_{\mu} W^c_{\nu} \right) \nonumber \\
{\cal O}_2^{(7)} &=& g' ( \overline{L^c} \bar{\epsilon} H ) \sigma^{\mu \nu} ( H^T
\bar{\epsilon} L )  \left(\partial_{\mu} B_{\nu} - \partial_{\nu} B_{\mu}  \right) 
\end{eqnarray}
and a correction to the neutrino mass would be generated by the operator 
\begin{equation}
{\cal O}_3^{(7)} = ( \overline{L^c} \bar{\epsilon} H ) ( H^T
\bar{\epsilon} L ) (H^{\dagger}H) \;\; .
\end{equation}
However, for Majorana neutrinos there is an even more stringent constraint imposed by the flavor symmetries of such neutrinos. Namely, Majorana neutrinos cannot have diagonal magnetic moments, only transition moments, either in the flavor or the mass basis, are possible. Hence the magnetic moment matrix in the flavor space is required to be antisymmetric in flavor indices even though the mass matrix is symmetric. This feature significantly weakens the constraints on the Majorana neutrino magnetic moments as compared to the Dirac case. In Ref.~\cite{Davidson:2005cs} it was shown that one-loop mixing of the mass and magnetic moment operators leads to rather weak constraints on the Majorana magnetic moment due to the suppression by charged lepton masses. Two-loop matching of the mass and magnetic field operators 
further reinforces this result \cite{Bell:2006wi}. The most general bound given in this reference is
\begin{equation}
|\mu_{\alpha \beta}| \le 4 \times 10^{-9} \mu_B \left( \frac{[m_{\nu}]_{\alpha \beta}}{1 \> {\rm eV}} \right) \left( \frac{1 \> {\rm TeV}}{\Lambda} \right)^2 
\frac{m_{\tau}^2}{|m_{\alpha}^2 -m_{\beta}^2|} \;\; .
\end{equation}

These arguments suggest that if the value of the neutrino magnetic moment is measured to be just below the present experimental and observational limits, then neutrinos are very likely Majorana fermions. 

\section{CONCLUSIONS}

Neutrinos are unique among all the elementary fermions of the Standard  Model: they carry no electric charges. This feature makes it possible for them to possess Majorana masses. There are very interesting consequences of this possibility. In this article we first reviewed the current status of our knowledge of neutrino properties, then explored theoretical motivations of experiments that can identify whether  neutrinos are Dirac or Majorana fermions. This experimental task is not easy since all the neutrinos that are directly observed are ultra relativistic. When they are ultra relativistic, Dirac and  Majorana neutrinos behave exactly the same way. Nevertheless, there are a handful of possibilities, which we elaborated on in some detail, ranging from neutrinoless double beta decay to the angular distribution of the decay products of heavy neutrinos. Numerous experiments exploring the neutrino properties are in progress or at the planning stage. The much-anticipated answer to the question of Dirac versus Majorana nature may not be too far away. 

\section*{DISCLOSURE STATEMENT}
The authors are not aware of any affiliations, memberships, funding, or financial holdings that
might be perceived as affecting the objectivity of this review. 

\section*{ACKNOWLEDGMENTS}
We thank Andr{\'e} de Gouv{\^e}a for many illuminating discussions. This work was supported in part by the US National Science Foundation Grant No. PHY-1514695 at the University of Wisconsin.
The document was prepared using the resources of the Fermi National Accelerator Laboratory (Fermilab), a U.S. Department of Energy, Office of Science, HEP User Facility. Fermilab is managed by Fermi Research Alliance, LLC (FRA), acting under Contract No. DE-AC02-07CH11359.

\end{document}